# El ciclo de las manchas solares y la precipitación en la Región del Eje Cafetero - Colombia

Fernando Alfonso González-Lozano

*Facultad de Ingeniería y Arquitectura, Universidad Nacional de Colombia, Manizales.*
*fagonzalez@unal.edu.co*

**Resumen**

Con el propósito de detectar la posible influencia del ciclo de la actividad solar en el clima regional, se establece mediante correlaciones, coeficientes de Pearson y Spearman, una relación inversa entre el índice de las manchas solares (*sunspot number*, *Observatory of Belgium*) y la precipitación anual en las estaciones: Cenicafé (1942-2014), Naranjal (1951-2014), La Bella (1951-2014), Miguel Valencia (1953-2011) ubicas en la Región del Eje Cafetero. Esta relación presenta una alta significación estadística para las tres primeras estaciones, mayor con la serie del índice de las manchas solares rezagada en uno y dos años. Las auto-correlaciones y los semivariogramas calculados hasta 24 rezagos para las series de precipitación anual de Cenicafé, Naranjal y La Bella, muestran de manera clara un ciclo de 10 o 11 años para las dos primeras estaciones, para La Bella se evidencia el ciclo en los rezagos con relación inversa, rezagos 6 y 16. El espectro de frecuencias de las precipitaciones de Cenicafé, Naranjal y La Bella, obtenido mediante la transformada rápida de Fourier, muestra resultados contradictorios, mientras en el espectro de frecuencias de Cenicafé apenas se insinúa el ciclo de los 11 años, el espectro de frecuencias de Naranjal privilegia un ciclo de 11 años, y el de La Bella muestra la fuerte presencia de ciclos de 11 y 22 años. Se expone una posible explicación de esta aparente contradicción. Finalmente, para comprobar un acoplamiento entre las precipitaciones regionales y el ciclo de 11 años de la actividad solar, se realizó una comparación estadística entre el promedio de precipitación de los años coincidentes con los mínimos en el ciclo de las manchas solares y el promedio de precipitación de los años coincidentes con los máximos de ese ciclo, la cual resultó con una alta significación estadística para las series de Cenicafé, Naranjal y La Bella. Corroborándose de esta forma la influencia del ciclo de las manchas solares en las precipitaciones regionales, lo cual es útil en el pronóstico a largo plazo de la disponibilidad hídrica regional.

*Palabras clave*: manchas solares, ciclos precipitación, Región del Eje Cafetero, cambio climático, auto-correlación.

**Abstract**

In order to detect a possible influence of the solar activity cycle on the regional climate, it is established by correlations, Pearson's and Spearman's coefficient, an inverse relation between the sunspots index (sunspot number, source: "WDC-SILSO, Royal Observatory of Belgium, Brussels") and the annual precipitation in the following stations: Cenicafé (1942-2014), Naranjal (1951-2014), La Bella (1951-2014), Miguel Valencia (1953-2014), localized in the Region of the Eje Cafetero and monitored by the Centro Nacional de Investigaciones del Café (Cenicafé). This relation evidences a high statistical significance for the first three stations being higher with the sunspot index series delayed one and two years. The autocorrelations and the semivariograms calculated until 24 delays for the Cenicafe, Naranjal and La Bella stations' series clearly show a cycle of 10 or 11 years for the first two stations. For La Bella station it is evident the cycle in the delays with inverse relation, delays 6 and 16. The frequency spectra of the Cenicafé, Naranjal and La Bella stations' precipitations, obtained by the rapid Fourier transform, show contradictory results, while the Cenicafé's frequency spectrum just denotes the cycle of 11 years, the Naranjal's frequency spectrum favors a cycle of 11 years, and La Bella shows the strong presence of cycles of 11 and 22 years. A possible explanation for this apparent contradiction is exposed. Finally, to prove a coupling between regional precipitation and the 11-year cycle of solar activity, it is performed a statistical comparison between the precipitation average of the coincident years with the minimum values of the sunspot cycle and the precipitation average of the coincident years with the peaks of this cycle. This comparison evidences a high statistical significance for the Cenicafe, Naranjal and La Bella stations' series. Demonstrating the influence of the sunspot cycle in the regional precipitations. This result is useful in the the long-term prevision of regional hydric availability.

## 1 Introducción

La influencia de la actividad solar en el clima de la tierra ha sido objeto de análisis, y en no pocos casos de controversias, entre los científicos y estudiosos del clima. Las manchas solares son un indicativo de las variaciones en la actividad solar, si bien la contribución de los ciclos de las manchas solares, 11 años, a la variación del balance de energía en la baja troposfera es de apenas 0,1%, unos $0.3 Wm^{-2}$ según mediciones satelitales [1], estudios recientes presenta evidencias crecientes de que la actividad solar tiene una influencia en el clima terrestre mediante una conexión entre la estratosfera y la troposfera como influencia de los ciclos solares [2, 3].

Algunos autores han postulado a las fluctuaciones de la incidencia sobre la tierra de los rayos cósmicos galácticos como una posible causa de variaciones climáticas, incidencia que es controlada por la actividad solar, cuando esta aumenta (disminuye) los rayos cósmicos incidentes disminuyen (aumentan). Los rayos cósmicos, a su vez, afectarían el clima terrestre propiciando la formación amplias coberturas de nubes bajas [1]. Mediante la concentración de radiocarbono $^{14}C$, una medida indirecta de la intensidad de los rayos cósmicos, se ha logrado establecer que durante período medieval cálido, 1000-1300 DC, cuando los vikingos se establecieron en Groenlandia, la actividad solar fue alta y la incidencia de rayos cósmicos baja. Luego la actividad solar decreció y la incidencia de rayos cósmicos aumento hasta llegar a los mínimos de Spörer 1460-1550 y el de Maunder 1645-1715, con la ausencia total de manchas solares reflejo de una baja actividad solar y también el de Dalton 1795-1825 con escasas manchas solares. El continente Euroasiático y Norteamérica, por lo menos, entran en un período frío de cerca de cuatro siglos 1450-1840 denominado la Pequeña Edad del Hielo, que cobija tanto el mínimo de Spörer, como el de Maunder y el de Dalton [1, 4, 5].

La Pequeña Edad de Hielo también se manifestó en el territorio colombiano, en general, y en el Eje Cafetero, en particular, como lo demuestran los hitos geomorfológicos en el parque Natural los Nevados, en especial las morrenas recientes de las laderas del Nevado de Santa Isabel, que se encuentran a los 4400 m de altitud [6, 7]. Las crónicas de los historiadores de Indias y algunos documentos coloniales sobre prácticas agrícolas del entonces, dan noticia de una época claramente más fría que los subsiguientes siglos XIX y XX [7].

Habida una posible conexión entre la actividad solar y el clima regional, cabe preguntarse: ¿afectan los ciclos de 11 años de las manchas solares y el de 22 años del campo magnético solar el clima de la Región del Eje Cafetero? Para contestar esta pregunta se recurre a la aplicación de métodos estadísticos: correlaciones, coeficientes de Pearson y Spearman, entre la serie del número de las manchas solares (sunspot number) y las series de precipitación anual de cuatro estaciones climatológicas operadas por el *Centro Nacional de Investigaciones del Café – Cenicafé-*, con más de 60 años de registros, ubicadas en la cuenca del Cauca medio; Cenicafé, Naranjal, La Bella, Miguel Valencia. Auto-correlación, con los coeficientes Pearson y Spearman, de la serie de precipitación de la estaciones climatológicas de Cenicafé, Naranjal y La Bella, semivariogramas de las mismas series, comparación estadística entre el promedio de precipitación de los años coincidentes con los mínimos en el ciclo de las manchas solares y el promedio de los años coincidentes con los máximos en ese ciclo y, finalmente, el análisis del espectro de frecuencias de la series de precipitación de Cenicafé, Naranjal y La Bella mediante la aplicación de la transformada rápida de Fourier.

La detección de un ciclo de baja frecuencia, 11 o 22 años, en la dinámica de las precipitaciones regionales se hace dispendioso por la variabilidad temporal y espacial que otros factores le imponen al clima en Colombia, velando el efecto que pueda tener el ciclo de la actividad solar. Factores que actúan en diferentes escalas como la Zona de Confluencia Intertropical (ZCIT), El Niño – Oscilación del Sur (ENSO), la Oscilación Cuasi – Bienal (QBO), la corriente superficial del Chocó, el cambio climático, los sistemas convectivos de mesoescala, la oscilación de Madden - Julian, las circulaciones valle-montaña, las características fisiográficas locales, el cambio en el uso de la tierra, que actúan en diferentes escalas y con distinta profundidad [8, 9, 7].

Además, del interés científico por desentrañar posibles acoplamientos entre el ciclo de las manchas solares y las precipitaciones regionales, el detectar un comportamiento recurrente en las lluvias asociado a la actividad solar con ciclos de 11 y 22 años sería de gran utilidad en el pronóstico a largo plazo de la disponibilidad hídrica regional.

1. **Datos**

a) Como índice de la actividad solar se utilizó el *sunspot number* total promedio anual ubicado en la fuente: *"WDC-SILSO, Royal Observatory of Belgium, Brussels"*, en su versión 2.0 vigente desde el 1 de julio del 2015.
b) Se recurrió a los datos de precipitación anuales y mensuales de cuatro estaciones climatológicas, operadas por *Cenicafé*, por su largo período de cubrimiento, más de 60 años, y la cuidadosa depuración y crítica de su información: Cenicafé 1942-2014 (5° 00' N, 75° 34' W, 1310 m), Naranjal 1951-2014 (4° 58' N, 75° 39' W, 1380 m), La Bella 1950-2014 (4° 30' N, 75° 40' W, 1450 m), Miguel Valencia 1953-2011 (5° 36' N, 75° 51' W, 1620 m). Todas ellas ubicadas en la cuenca del Cauca Medio en la Región del Eje Cafetero, zona cafetera colombiana por excelencia, las tres primeras en las estribaciones de la Cordillera Central vertiente occidental y la última en la vertiente oriental de la Cordillera Occidental. Información pluviométrica publicada en el *Anuario Meteorológico Cafetero.*
*c)* Índice mensual de la anomalía de la temperatura superficial del mar (SST) en el paralelogramo Niño3 (5°N - 5°S, 150°W – 90°W). Fuente: *National Oceanic and Atmospheric Administration –NOAA-, USA.* Se escogió la anomalía SST Niño3, como índice del ENSO, por ser esta anomalía la de mayor correlación con las series pluviométricas regionales y porque este paralelogramo es adyacente a la Costa Pacífica colombiana [7].

2. **Métodos**

Se aplicaron métodos de la estadística descriptiva convencional como son: los promedios mensuales multianuales, la correlación entre series y la auto-correlación, con base en los estadísticos de Pearson y Spearman, y la comparación clásica entre los promedios de dos poblaciones. Además, también se aplicó el método descriptivo del semivariograma y el método de análisis espectral con base en la transformada rápida de Fourier. Métodos que fueron aplicados haciendo uso del paquete estadístico *IBM SPSS Statistics versión 22,* excepto el semivariograma que se programó*,* y los resultados dibujados en figuras mediante la hoja de cálculo *Excel Office,* a continuación de describen de manera sucinta:

- Para caracterizar el régimen pluviométrico multianual e ilustrar la magnitud de las lluvias mensuales, se calcularon los promedios mensuales multianuales de las cuatro estaciones climatológicas con los cuales se construyó una figura ilustrativa.

- Correlaciones, coeficientes de Pearson y Spearman, entre las series de precipitación anual de las cuatro estaciones y la serie de las manchas solares promedio anuales (sunspot number) sus primeros tres rezagos y la serie de la anomalía de la temperatura superficial del mar en Niño3 y su primer rezago.

El coeficiente de correlación de Pearson se simboliza con la letra $\rho_{X,Y}$ siendo su expresión matemática:

$$\rho_{X,Y} = \frac{\sigma_{XY}}{\sigma_X \sigma_Y} = \frac{E[(X-\mu_X)(Y-\mu_Y)]}{\sigma_X \sigma_Y}$$

donde: $\sigma_{XY}$ es la covarianza de $(X, Y)$
$\sigma_X$ es la desviación típica de la variable $X$
$\sigma_Y$ es la desviación típica de la variable $Y$

El coeficiente de correlación de Spearman es una medida no-paramétrica de la asociación entre dos variables, que cuando la función es monótona creciente o decreciente infiere una correlación perfecta. En contraste con el coeficiente de Pearson que infiere correlación perfecta cuando la función es lineal. El coeficiente de Spearman es menos sensible a valores atípicos y logra detectar relaciones no-lineales que el coeficiente de Pearson desecharía. Parte de ordenar las dos variables; $X$, $Y$ y para cada par medir la distancia entre la ubicación de *xi* y *yi*, de ahí su nombre en inglés: *Spearman's rank correlation coefficient*. Su expresión matemática:

$$\rho = 1 - \frac{6 \sum d_i^2}{n(n^2 - 1)}$$

Donde: $d_i$ es la diferencia en el orden entre *xi* y *yi*
$n$ es el número de parejas

- Auto-correlación de la series de precipitación anual de las estaciones climatológicas de Cenicafé, Naranjal y la Bella con sus 24 primeros rezagos, haciendo uso de los coeficientes de Pearson y Spearman.

- Semivariograma de las series de precipitación anual de las estaciones de Cenicafé, Naranjal y La Bella. Función de distancia utilizada en la interpolación espacial, pero que aquí se hace uso de ella, para medir similitudes no en el espacio sino en el tiempo: de la serie y sus rezagos, tal como lo han planteado Peña, Jaramillo y Paternina [10], quienes describen en detalle el método. Su expresión matemática:

$$\gamma(h) = \frac{\sum (Z(t+h) - Z(t))^2}{2n}$$

Donde: $\gamma(h)$ semivarianza
$h$ rezago en años
$Z(t)$ valor de la función
$n$ cantidad de pares separados un tiempo $h$

El semivariograma se gráfica dividiendo la semivarianza por la varianza muestral. Su interpretación es distinta a la de la correlación: valores inferiores a uno indican alta similitud de la serie y su rezago, valores alrededor de uno muestran que no existe relación entre la serie y su rezago, y valores altos, superiores a uno, una relación inversa entre la serie y su rezago. Este método se programó ya que el paquete estadístico *SPSS* no cuenta con él.

- Espectro de frecuencia de las series anuales de precipitación de Cenicafé, Naranjal y La Bella, índice de las manchas solares (*sunspot number*) y anomalías SST en Niño3. El análisis espectral de una señal digital consiste en identificar sus distintos componentes dentro del dominio de la frecuencia, lo cual se obtiene mediante la transformada rápida de Fourier:

$$X_k = \sum_{n=0}^{N-1} x_n \, e^{-\frac{2\pi i}{N}kn} \qquad k = 0, \ldots, N-1$$

En las ecuaciones los números complejos $X_k$ están representando la amplitud y fase de diferentes componentes sinusoidales de la señal $x_n$, es decir, de la serie.

- Comparación entre medias. La prueba de diferencia entre dos muestras se utiliza para decidir si las medias de dos poblaciones son iguales. Se definen así las dos poblaciones para cada una de las cuatro estaciones : Cenicafé, Naranjal, La Bella y Miguel Valencia:
    a) las precipitaciones de los años coincidentes con los mínimos solares y de los dos años subsiguientes.
    b) las precipitaciones de los años coincidentes con los máximos solares y de los dos años subsiguientes.

Se plantea la hipótesis nula *H0: μx = μY* , o su equivalentemente, *H0: μx - μY = 0* . El estadístico *t* para probar si las medias poblacionales son distintas puede ser calculado como sigue:

$$t = \frac{\bar{X}-\bar{Y}}{S_{\bar{X}-\bar{Y}}} \qquad S_{\bar{X}-\bar{Y}} = \sqrt{\frac{S_x^{\,2}}{n_x}+\frac{S_y^{\,2}}{n_y}}$$

Donde:  $n_x$ , $n_y$ tamaño de las muestras
$\bar{X}$ , $\bar{Y}$  estimadores muestrales de las medias
$S_x^{\,2}$, $S_y^{\,2}$  estimadores muestrales de las varianzas

## 3. Resultados

- El patrón o régimen pluviométrico multianual de las cuatro estaciones analizadas se caracterizan por una forma de doble onda, "bimodal", de la distribución de las lluvias en el transcurso del año, típica del ecuador climático, cuyos máximos coinciden con el mayor calentamiento atmosférico que sucede en los días posteriores a las dos posiciones cenitales del sol sobre la región. El patrón pluviométrico de doble onda típico del ecuador climático de la Región del Eje Cafetero, ocasionado por el doble paso de la ZCIT, sufre variaciones según prevalezca la influencia de los Alisios del Norte o de los Alisios del Sur. En el norte de la Región, en Miguel Valencia (5° 36' N), se sucede un período de relativa sequía más pronunciado a final del año, en el trimestre DEF. En el Sur, en La Bella (4° 30' N), ocurre a la inversa el período de relativa sequía más pronunciado se sucede a mitad del año, en el trimestre JJA. En las estaciones Cenicafé (5° 00' N) y Naranjal (4° 58' N) el régimen pluviométrico multianual es enteramente "bimodal", indicativo de la ubicación geográfica del ecuador climático [7, 11]. La Figura 1 muestra lo anterior e ilustra la magnitud de las lluvias mensuales.

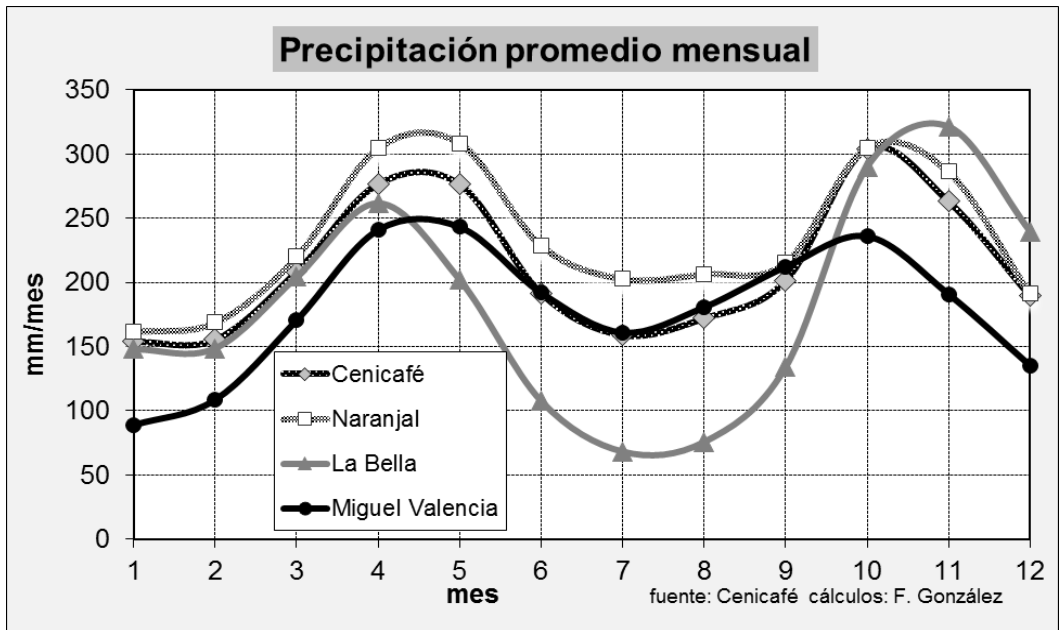

Figura 1. Regímenes de precipitación Región del Eje Cafetero.

- La Tabla 1 muestra las correlaciones, coeficientes de Pearson y Spearman, entre las series de precipitación anual de las cuatro estaciones, el índice de las manchas solares (sunspot number –SSN-) y sus tres primeros rezagos, y la serie de la anomalía de la temperatura superficial en Niño3 y su primer rezago.

Las correlaciones entre las series pluviométricas son muy altas y estadísticamente significativas al 99%. La mayor, 0,870, se da entre Cenicafé y Naranjal estaciones cercanas, y la menor, 0,593, entre las estaciones más alejadas: La Bella y Miguel Valencia.

Las series pluviométricas están estrechamente relacionadas con SSN y sus dos primeros rezagos, salvo la serie de Miguel Valencia que sólo tuvo una correlación significativa al 95% con el segundo rezago calculada con el coeficiente de Spearman. Las correlaciones de las series de Cenicafé, Naranjal y La Bella con las series del SSN y sus rezagos son estadísticamente significativas, en especial, con el primer y segundo rezago con significación del 99%. Las correlaciones con el tercer rezago no poseen significación estadística, excepto la de Naranjal.

| TABLA 1 Correlaciones Precipitación Anual - Sun Spot Number - Anomalía SST Niño3 (coef. Pearson y Spearman) | | | | | | | | | | | |
|---|---|---|---|---|---|---|---|---|---|---|---|
| | | | Precipitaciones anuales | | | Sun Spot índice y rezagos | | | | Anomalía SST Niño3 rezago | |
| | | | PNaranjal | PLaBella | PMiguelVal | SunSpot | SunSpot(-1) | SunSpot(-2) | SunSpot(-3) | AnomNiño3 | AnomNiño3(-1) |
| PCenicafé | Pearson | coeficiente | 0.867 | 0.724 | 0.729 | -0.317 | -0.374 | -0.349 | -0.187 | -0.505 | -0.038 |
| | | test p.valor | 0.000 | 0.000 | 0.000 | 0.006 | 0.001 | 0.003 | 0.113 | 0.000 | 0.764 |
| | Spearman | coeficiente | 0.861 | 0.636 | 0.732 | -0.318 | -0.377 | -0.351 | -0.173 | -0.490 | -0.013 |
| | | test p.valor | 0.000 | 0.000 | 0.000 | 0.006 | 0.001 | 0.002 | 0.144 | 0.000 | 0.919 |
| | n | año | 64 | 64 | 59 | 73 | 73 | 73 | 73 | 65 | 64 |
| PNaranjal | Pearson | coeficiente | 1 | 0.682 | 0.699 | -0.328 | -0.435 | -0.408 | -0.266 | -0.412 | -0.025 |
| | | test p.valor | | 0.000 | 0.000 | 0.008 | 0.000 | 0.001 | 0.033 | 0.001 | 0.845 |
| | Spearman | coeficiente | 1 | 0.607 | 0.691 | -0.318 | -0.444 | -0.444 | -0.280 | -0.400 | -0.018 |
| | | test p.valor | | 0.000 | 0.000 | 0.011 | 0.000 | 0.000 | 0.025 | 0.001 | 0.887 |
| | n | año | | 64 | 59 | 64 | 64 | 64 | 64 | 64 | 64 |
| PLaBella | Pearson | coeficiente | | 1 | 0.593 | -0.319 | -0.417 | -0.334 | -0.211 | -0.452 | -0.101 |
| | | test p.valor | | | 0.000 | 0.010 | 0.001 | 0.007 | 0.094 | 0.000 | 0.429 |
| | Spearman | coeficiente | | 1 | 0.609 | -0.337 | -0.413 | -0.320 | -0.179 | -0.495 | -0.046 |
| | | test p.valor | | | 0.000 | 0.006 | 0.001 | 0.010 | 0.157 | 0.000 | 0.716 |
| | n | año | | | 59 | 64 | 64 | 64 | 64 | 64 | 64 |
| PMiguelVal | Pearson | coeficiente | | | 1 | -0.131 | -0.228 | -0.248 | -0.189 | -0.538 | 0.161 |
| | | test p.valor | | | | 0.324 | 0.083 | 0.058 | 0.153 | 0.000 | 0.223 |
| | Spearman | coeficiente | | | 1 | -0.114 | -0.197 | -0.270 | -0.190 | -0.531 | 0.172 |
| | | test p.valor | | | | 0.388 | 0.135 | 0.038 | 0.150 | 0.000 | 0.194 |
| | n | año | | | | 59 | 59 | 59 | 59 | 59 | 59 |
| SunSpot | Pearson | coeficiente | | | | 1 | 0.799 | 0.370 | -0.126 | 0.065 | |
| | | test p.valor | | | | | 0.000 | 0.001 | 0.289 | 0.609 | |
| | Spearman | coeficiente | | | | 1 | 0.794 | 0.355 | -0.133 | 0.078 | |
| | | test p.valor | | | | | 0.000 | 0.002 | 0.263 | 0.539 | |
| | n | año | | | | | 75 | 74 | 73 | 65 | |
| AnomNiño3 | Pearson | coeficiente | | | | | 0.118 | 0.141 | 0.055 | 1 | 0.083 |
| | | test p.valor | | | | | 0.350 | 0.262 | 0.651 | | 0.514 |
| | Spearman | coeficiente | | | | | 0.114 | 0.160 | 0.079 | 1 | 0.087 |
| | | test p.valor | | | | | 0.365 | 0.203 | 0.529 | | 0.496 |
| | n | año | | calculos: | F. González | | 65 | 65 | 65 | | 64 |

La serie de Miguel Valencia sólo cuenta con 59 años. Se comprobó que las correlaciones de las series de las otras estaciones con el SSN y sus dos primeros rezagos calculadas en el mismo período de cubrimiento que el de la estación Miguel Valencia no sufren cambios que afecten su significación estadística. Así pues, no es por poseer un período de cubrimiento menor que Miguel Valencia presenta un comportamiento diferente a las otras estaciones con respecto a la relación con el SSN.

La relación entre las precipitaciones regionales y el ENSO es estrecha, como lo comprueban las altas correlaciones negativas entre las series pluviométricas y las anomalías SST en Niño3. Hecho que lo han manifestado numerosos estudios para Colombia, en general, y para el Eje en Cafetero, en particular [8, 7]. No hay relación entre las precipitaciones regionales y la serie de la anomalía SST Niño3 rezagada.

La correlación de SSN con su primer rezago es muy alta y aún tiene alta significación estadística con su segundo rezago. No así la anomalía SST Niño3 con su inmediato rezago. Lo que tiene importancia al construir modelos de pronóstico de lluvia a un plazo de años.

Las correlaciones entre las anomalías SST Niño3 y el SSN y sus rezagos son bajas y estadísticamente despreciables, lo que indica la baja o nula relación entre el ENSO y el ciclo solar. Algunos estudios han llegado a la misma conclusión empleando métodos mucho más elaborados [12].

Con miras a tener una cuantificación que permita comparar el efecto del ciclo de las manchas solares y del ENSO en las lluvias regionales se construyó la Tabla 2. Donde se advierte, en el $R^2$ del modelo de regresión lineal univariado, que la influencia del ENSO es mayor que la del ciclo solar. Pero en tres estaciones climatológicas: Cenicafé, Naranjal y La Bella el efecto del ciclo solar no es nada despreciable, incluso en Naranjal es superior al del ENSO.

| TABLA 2 Variación explicada modelo de regresión lineal $R^2$ | | | | |
|---|---|---|---|---|
| | | SunSpot(-1) | SunSpot(-2) | AnomNiño3 |
| PCenicafé | $R^2$ (%) | 14,0 | 12,2 | 25,5 |
| | ANOVA p.valor | si | si | si |
| PNaranjal | $R^2$ (%) | 18,9 | 16,7 | 17,0 |
| | ANOVA p.valor | si | si | si |
| PLaBella | $R^2$ (%) | 17,1 | 11,2 | 20,4 |
| | ANOVA p.valor | si | si | si |
| PMiguelValencia | $R^2$ (%) | 5,2 | 6,2 | 28,9 |
| | ANOVA p.valor | no | no | si |
| si: estadísticamente significativo al nivel del 99% de confianza | | | | |
| no: estadísticamente NO significativo al nivel del 95% de confianza | | | | |

- Los resultados anteriores advierten de un posible acoplamiento entre las variaciones de las manchas solares y las lluvias regionales. Cabe preguntarse entonces; si el ciclo solar de 11 años se manifiesta en la dinámica de las precipitaciones. Para lo cual se calcularon las auto-correlaciones, coeficientes de Pearson y Spearman, de las series de precipitación anual de Cenicafé, Naranjal y La Bella que presentaron correlaciones significativas con el SSN. Lo que se muestra en las figuras 2, 3 y 4.

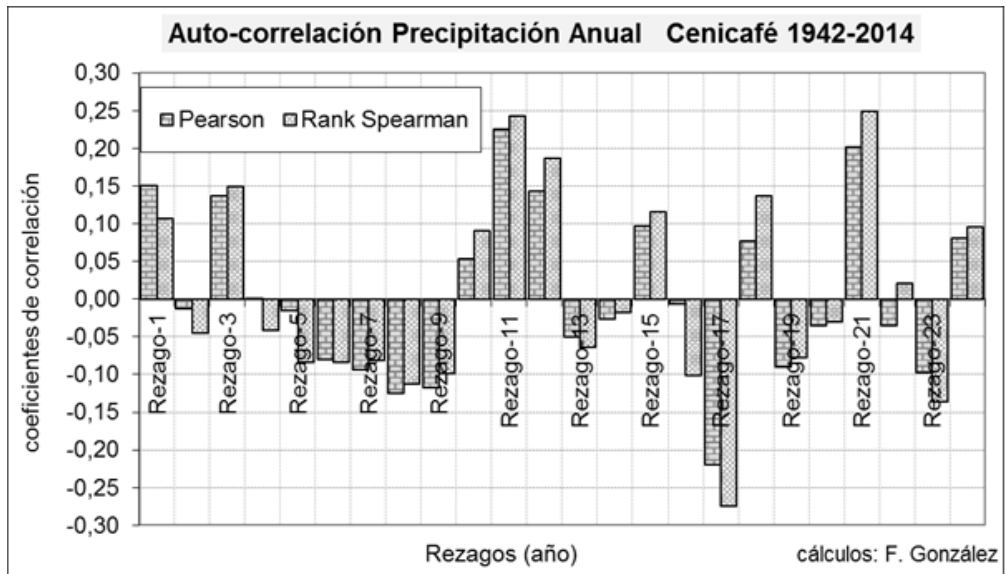
Figura 2. Auto-correlación, serie precipitación anual Cenicafé (24 rezagos).

La Figura 2 advierte que las mayores correlaciones positivas se encuentran con los rezagos de orden 11 y 21. Con el regazo 17, al estar en la fase inversa del ciclo, los valores de la correlación son negativos, la correlación de este rezago calculada con el coeficiente de Spearman es estadísticamente significativa al nivel del 95%.

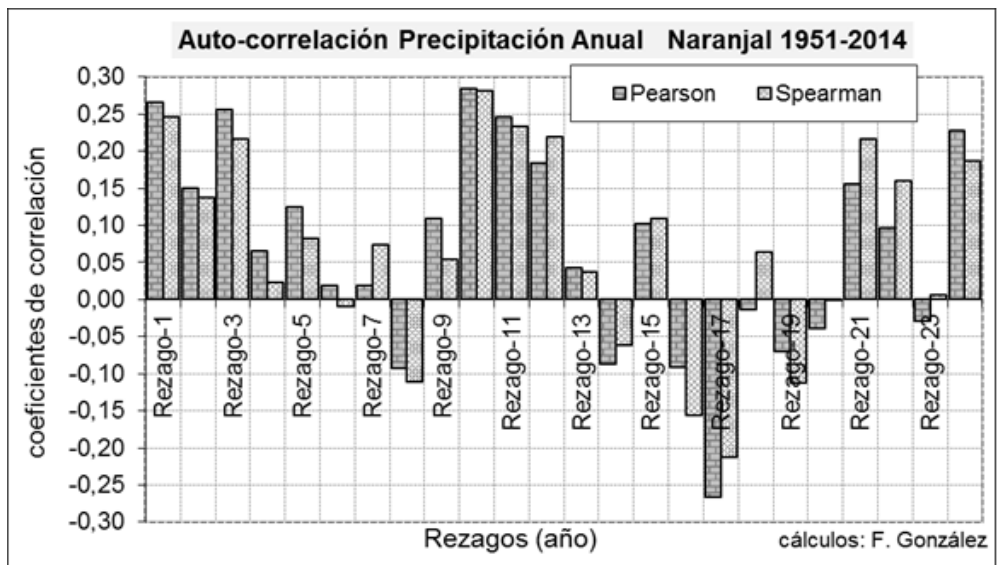
Figura 3. Auto-correlación, serie precipitación anual Naranjal (24 rezagos).

Las auto-correlación de Naranjal, Figura 3, es estadísticamente significativa al 95% con el coeficiente de Pearson para los rezagos 1, 3 y 10, para el rezago 10, también es estadísticamente significativa con el coeficiente de Spearman, evidencia de un posible ciclo decadal.

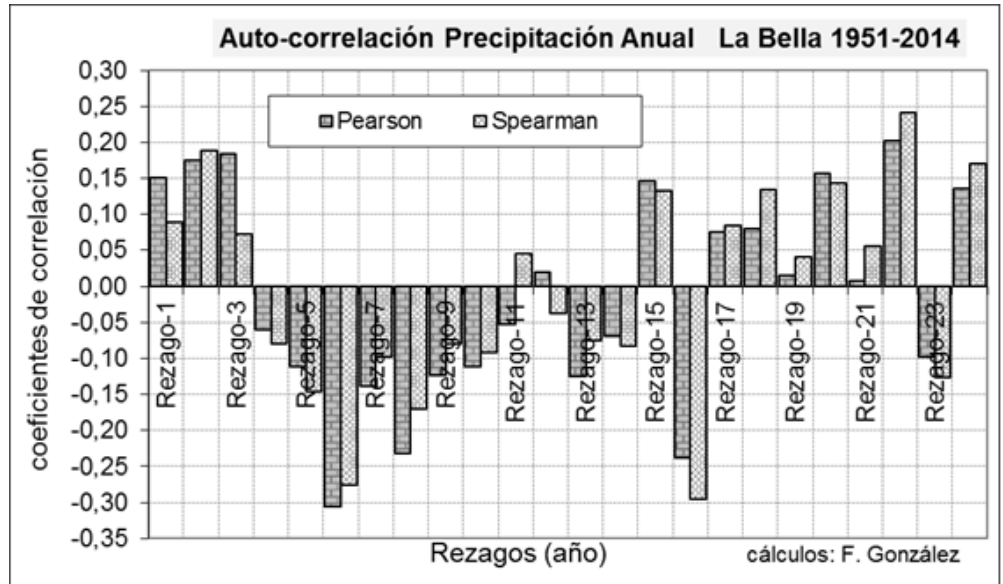

Figura 4. Auto-correlación, serie precipitación anual La Bella (24 rezagos).

La Figura 4 muestra correlaciones negativas altas en los rezagos 6 y 16 que se encuentran en la fase inversa del ciclo, correlaciones que son estadísticamente significativas al 95% de confianza.

La auto-correlaciones de la precipitación anual en las tres estaciones muestran correlaciones altas en rezagos alrededor del ciclo decadal, rezagos 10 y 11, o en la fase inversa del ciclo, rezagos 6, 16 y 17, que alcanzan a tener significación estadística. Lo anterior no es prueba amplia y suficiente de un ciclo de 11 años en las precipitaciones regionales, pero si es un indicio importante de la existencia del ciclo. Por lo tanto, se procede a hacer algunos otros análisis.

- Recientemente se ha aplicado la función de distancia denominada semivariograma para detectar ciclos de baja frecuencia en las precipitaciones regionales [10], aquí se hace uso de la misma técnica descriptiva con las series de precipitación anual de Cenicafé, Naranjal y La Bella, pero sin la suavización de promedios móviles que allí se aplicó. La Figura 5 muestra el semivariograma de las tres series de precipitación (semivarianza / varianza muestral).

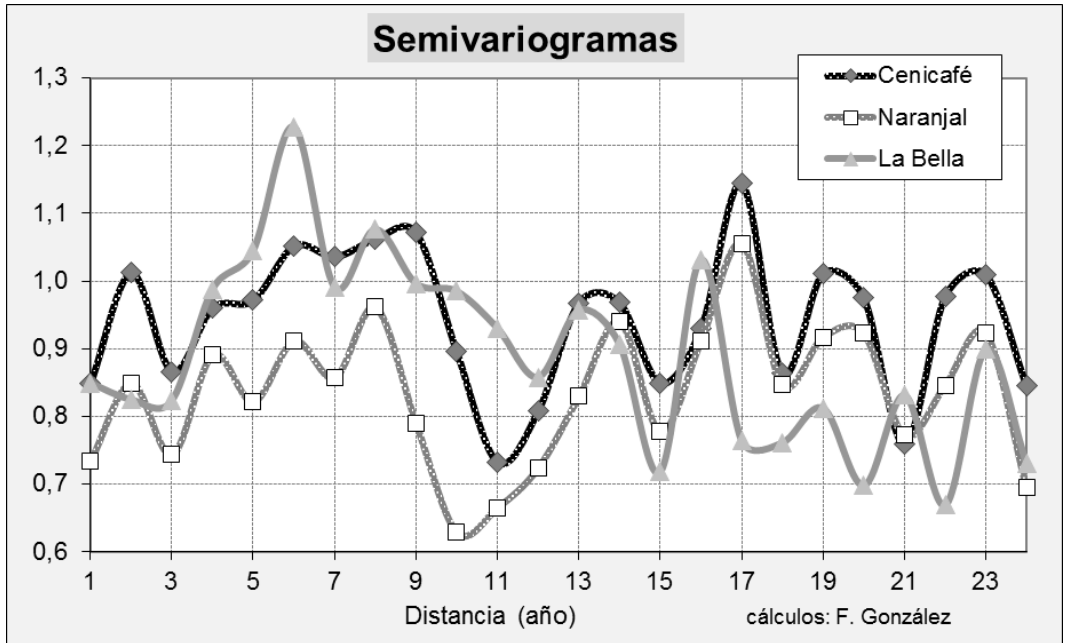
Figura 5. Semivariogramas Precipitación anual de Cenicafé, Naranjal y La Bella (semivarianza/varianza muestral).

Los semivariogramas de las tres series, Figura 5, corroboran, *sensu lato*, lo hallado en las auto-correlaciones. Advierte acercamientos notorios en los rezagos 10 y 11 en la serie de Naranjal y en el rezago 22 de la serie de La Bella. Alejamientos notorios en el rezago 6 de la serie de La Bella y en el rezago 17 de Cenicafé, rezagos en la fase inversa del ciclo. Recuérdese que el semivariograma es una función de distancia, entre más cercana a cero, mayor es la similitud de la serie con su rezago, valores superiores a uno indican una relación inversa. Resultado que evidencia la existencia de un ciclo de 11 años, que la auto-correlación ya lo había mostrado.

- Una técnica matemática para identificar densidades de frecuencias asociadas a periodos de recurrencia es el análisis espectral de una señal digital, las cuales se obtiene mediante la transformada rápida de Fourier. Análisis provisto por el paquete *SPSS*. Se aplicó la técnica a las series de precipitaciones anuales de Cenicafé, Naranjal, La Bella, el SSN (sunspot number), y la anomalía SST en Niño3 (AnomNiño3). Las figuras 6, 7, 8, 9 y 10 sintetizan los resultados.

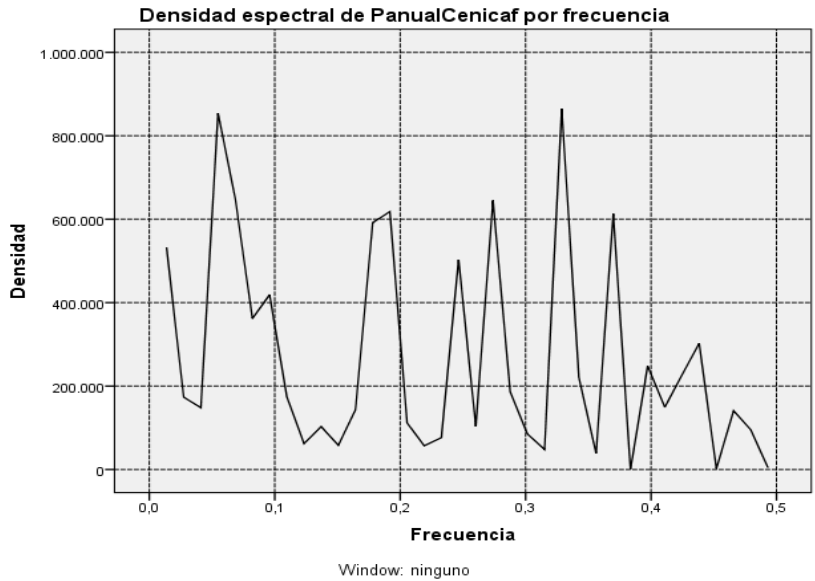

Figura 6. Densidad espectral por frecuencia Panual Cenicafé

El espectro de frecuencias de Cenicafé privilegia un ciclo de alrededor de los 20 años y otro de aproximadamente tres años.

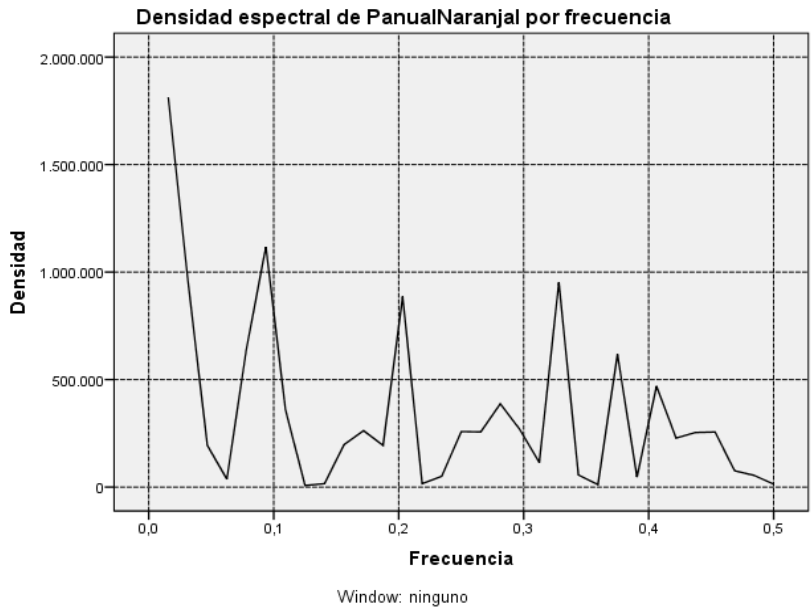

Figura 7. Densidad espectral por frecuencia Panual Naranjal

La densidad espectral de Naranjal muestra como el ciclo de mayor relevancia el de los 11 años. También destaca un ciclo de aproximadamente tres años y otro de cerca de cinco años.

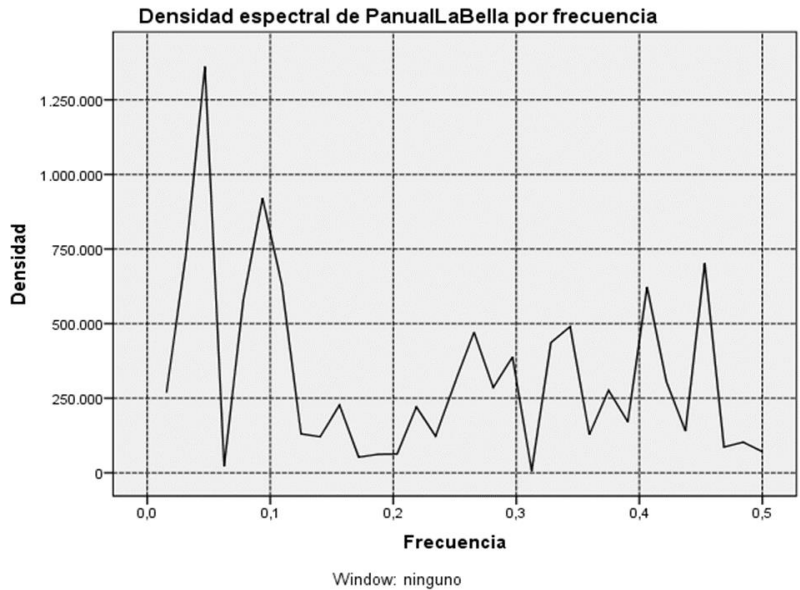

Figura 8. Densidad espectral por frecuencia Panual La Bella

La densidad espectral de La Bella mostrada en la Figura 8 privilegia, de manera notoria, dos ciclos: uno de aproximadamente 22 años y otro alrededor de 11 años.

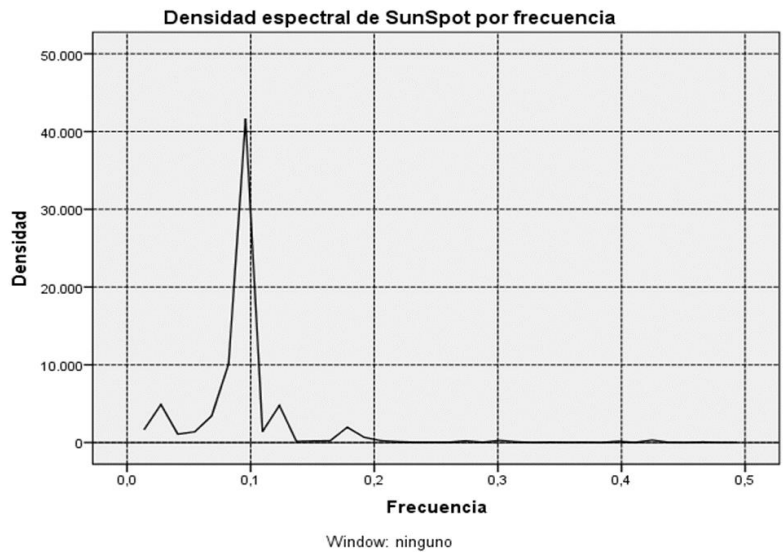

Figura 9. Densidad espectral por frecuencia SSN

La Figura 9 de la densidad espectral del SSN, advierte, de manera notable, de su ciclo de 11 años.

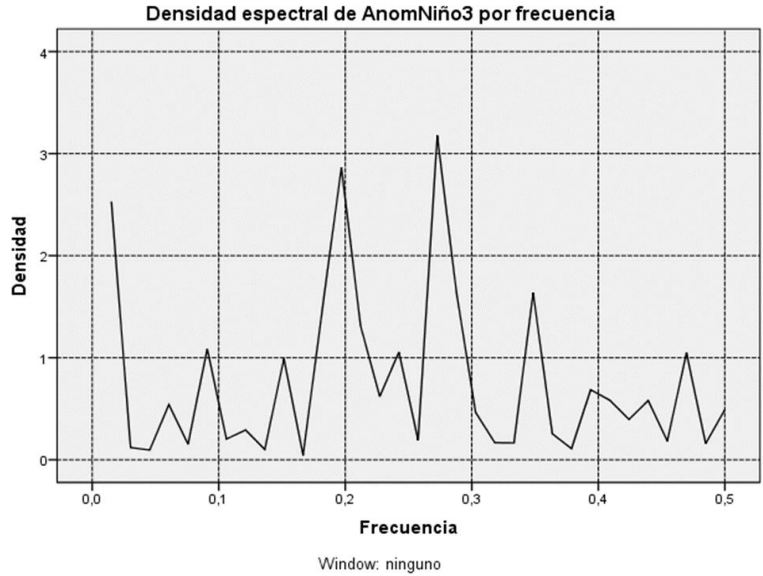

Figura 10. Densidad espectral por frecuencia AnomNiño3

La densidad espectral de AnomNiño3, Figura 10, privilegia dos ciclos: uno de unos 3,7 años y otro de unos 5,1 años.

En conclusión, la densidad espectral de las series de precipitación anual regionales, no "heredan" las propiedades de la densidad espectral de AnomNiño3. Las series de Naranjal y La Bella, muestran el ciclo de 11 años coincidente con el ciclo del SSN.

- Por último, para corroborar la influencia del ciclo de las manchas solares en las lluvias aforadas en Cenicafé, Naranjal, La Bella y Miguel Valencia y cuantificar su efecto, se tomaron dos poblaciones de precipitaciones anuales, considerando la correlación inversa estadísticamente significativa de las precipitaciones regionales con el SSN y sus dos primeros rezagos:
    a) P.Solmin; las precipitaciones de los años coincidentes con los mínimos solares y de los dos años subsiguientes, marcadas con un cuadrado, (■), en la Figura 11.
    b) P.Solmax; las precipitaciones de los años coincidentes con los máximos solares y de los dos años subsiguientes, marcadas con un triángulo, (▲), en la Figura 11.

El máximo del año 1957 en la serie SSN se extendió a los años de 1958 y 1959 por sus valores extremos, los mayores de la serie. De manera similar, el mínimo profundo del año 2007 se extendió a los años 2008 y 2009, los menores valores de la serie.

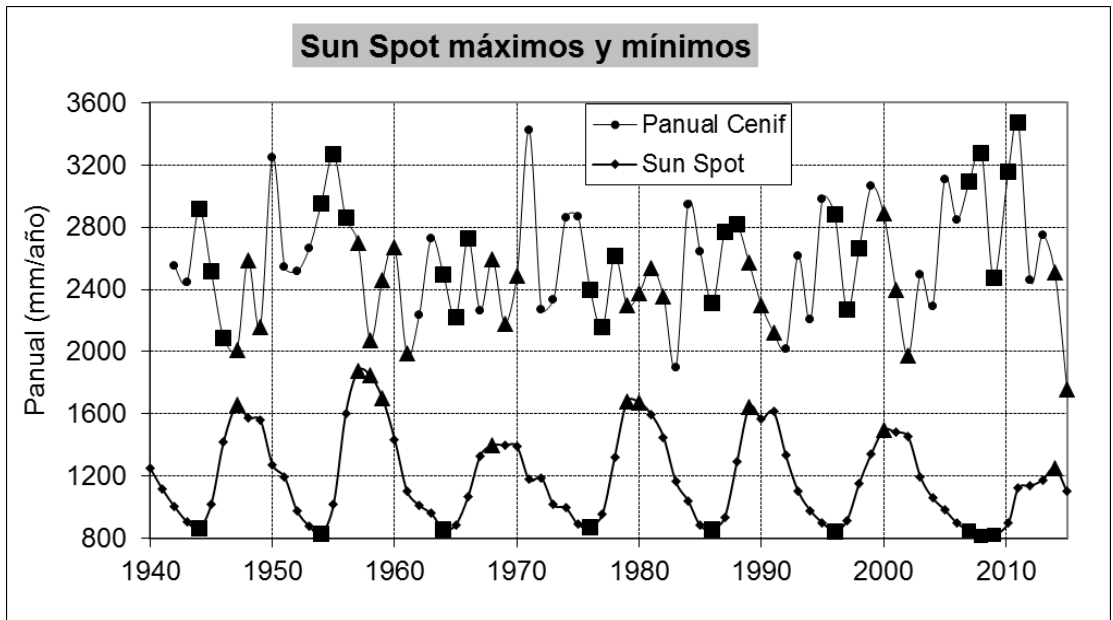

Figura 11. Series de precipitación Cenicafé (arriba) y SSN (abajo) con sus máximos (▲) y mínimos señalados (■).

La Figura 12, gráfico de box - whisker muestra estadísticos de las dos poblaciones de precipitación anual en Cenicafé (serie con el período de cubrimiento más largo, 73 años): valor mínimo, cuartil inferior, media, mediana, cuartil superior, valor máximo. Nótese que tanto la media como la mediana de P.Solmax están por debajo del cuartil inferior de P.Solmin, el cuartil superior de P.Solmax está muy por debajo de la media y la mediana de P.Solmin, el máximo de P.Solmax no alcanza el tercer cuartil de P.Solmin.

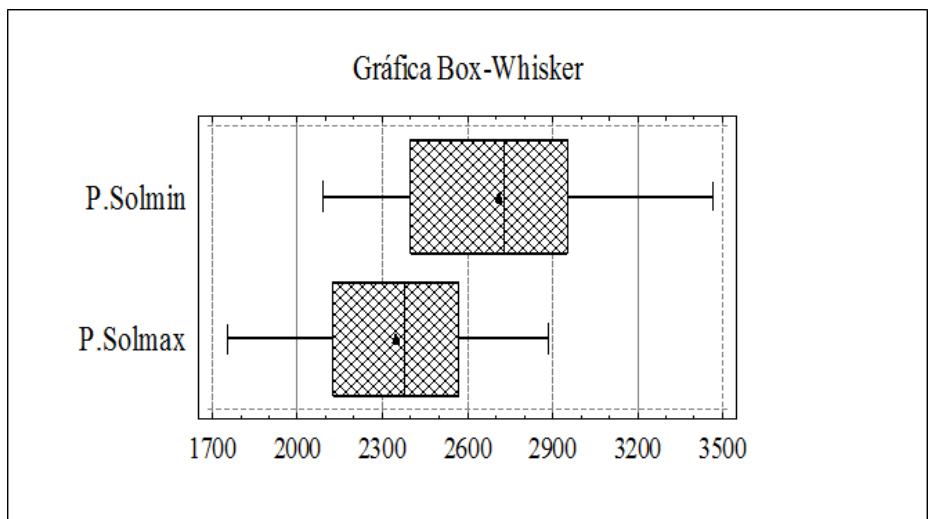

Figura 12. Cenicafé, P.Solmin y P.Solmax gráfico box - whisker

A esas dos poblaciones: P.Solmin y P.Solmax de precipitaciones anuales de las cuatro series de precipitación: Cenicafé, Naranjal, La Bella y Miguel Valencia, se les aplicó la prueba de diferencia de medias. Los resultados se muestran en la Tabla 3, nótese el intervalo del 99% de confianza estrictamente positivo de la diferencia de medias de las estaciones: Cenicafé, Naranjal y La Bella, es decir, que las poblaciones tienen medias diferentes desde el punto de vista estadístico. No así la estación Miguel Valencia, que al 95% de confianza, las medias de P.Solmin y P.SolMax no difieren.

| TABLA 3 Prueba de diferencia de medias entre P.Solmin y P.Solmax | | | | | | intervalo de confianza del 95% | | intervalo de confianza del 99% | |
|---|---|---|---|---|---|---|---|---|---|
| | | | | (mm/año) | | | | | |
| estaciones | | n (año) | Media | Desviación estándar | Media del error estándar | Inferior | Superior | Inferior | Superior |
| **Cenicafé** | P.Solmin | 23 | 2712 | 385 | 80 | 2546 | 2879 | 2486 | 2938 |
| | P.Solmax | 23 | 2348 | 279 | 58 | 2227 | 2469 | 2184 | 2512 |
| | Diferencia | | 364 | | 99 | 164 | 565 | 96 | 632 |
| | de medias | Se concluye que las medias son diferentes al 99% de confianza | | | | | | | |
| **Naranjal** | P.Solmin | 20 | 2974 | 444 | 99 | 2766 | 3182 | 2690 | 3258 |
| | P.Solmax | 20 | 2550 | 332 | 74 | 2394 | 2705 | 2337 | 2762 |
| | Diferencia | | 424 | | 124 | 173 | 676 | 87 | 762 |
| | de medias | Se concluye que las medias son diferentes al 99% de confianza | | | | | | | |
| **La Bella** | P.Solmin | 20 | 2360 | 495 | 111 | 2128 | 2592 | 2043 | 2677 |
| | P.Solmax | 20 | 1954 | 284 | 63 | 1821 | 2087 | 1773 | 2136 |
| | Diferencia | | 406 | | 128 | 145 | 666 | 55 | 757 |
| | de medias | Se concluye que las medias son diferentes al 99% de confianza | | | | | | | |
| **Miguel Valencia** | P.Solmin | 20 | 2238 | 410 | 92 | 2046 | 2430 | 1975 | 2500 |
| | P.Solmax | 18 | 2050 | 224 | 53 | 1939 | 2161 | 1897 | 2203 |
| | Diferencia | | 188 | | 106 | -28 | 404 | -103 | 479 |
| | de medias | **No** se concluye que las medias son diferentes al 95% de confianza | | | | | | | |

El promedio de las precipitaciones anuales de Cenicafé coincidentes con los mínimos SSN y dos años subsiguientes, supera en un 15,5%, al promedio de las precipitaciones coincidentes con los máximos SSN y dos años subsiguientes. En Naranjal esa diferencia es del 16,6%, en La Bella 20,8% y en Miguel Valencia apenas es del 9,2%.

### 4. Conclusiones y discusión

- Las correlaciones inversas estadísticamente muy significativas del sunspot number (SSN) con las precipitaciones anuales en las estaciones de Cenicafé, Naranjal y La Bella, la auto-correlación y el semivariograma de las tres series que evidencian un ciclo decadal, las densidades espectrales de Naranjal y La Bella que muestran de manera clara el ciclo de 11 años, la diferencia de medias con alta significación estadística entre las precipitaciones en las tres estaciones de los años coincidentes con los mínimos de las manchas solares y los dos años subsiguientes (P.Solmin) y las precipitaciones coincidentes con los máximos y los dos años subsiguientes (P.Solmax), advierten de una modulación importante de las lluvias regionales por el ciclo de las manchas solares. Cuando el ciclo SSN está en su mínimo las precipitaciones tienden a ser mayores que cuando el ciclo está en su máximo.

- Es notable la coincidencia de la auto-correlación y el semivariograma de las precipitaciones anuales en Cenicafé, Naranjal y La Bella (figuras 2, 3, 4 y 5). En ellas se aprecia similitudes de las series con sus rezagos de orden decadal o bidecadal, rezagos 10, 11, 21 o 22, y alejamientos en la fase inversa del ciclo, rezagos 6, 16 o 17. Hecho que unido al ciclo de 11 años que muestran las densidades espectrales de Naranjal y La Bella, dan fe inequívoca de un comportamiento cíclico cercano a los 11 años de las precipitaciones regionales.

- La no aparición, en la gráfica de la densidad espectral de las precipitaciones de Cenicafé, del ciclo de 11 años puede deberse a la influencia del ENSO, aún mayor que la del SSN, según el valor de las correlaciones y la varianza explicada por el modelo de regresión lineal univariado (tablas 1 y 2). La densidad espectral de las anomalías en la temperatura superficial del mar en Niño3, muestra varios posibles ciclos, los más destacados de 3,1 y 5,7 años. Que tampoco se manifiestan en la densidad espectral de las lluvias en Cenicafé. Artículos de la literatura especializada hablan de los límites de los métodos espectrales en la detección de frecuencias de oscilación, y proveen ejemplos de series con idénticos periodogramas pero distintas frecuencias de oscilación [13, 14]. Algunos autores hablan de una recurrencia caótica del ENSO, entre 2 y 10 años, no de ciclos o variaciones periódicas [15].

- Si bien, según los análisis hechos, existe un efecto del ciclo solar en el clima de la Región del Eje Cafetero, en la estación Miguel Valencia, en el norte de la Región, el efecto es mínimo. Lo que indica que esta influencia no se extiende a toda la zona andina colombiana. ¿Que tanto se extiende en el espacio el acoplamiento entre manchas solares y precipitación? Para contestar esta pregunta se hace necesario estudiar otras series de estaciones pluviométricas ubicadas en otros lugares. Pero deben cumplir la condición de tener un período extenso de cubrimiento, al menos 70 años continuos, para cubrir seis ciclos de las manchas solares, buena calidad en su información, y su vecindario no haya tenido modificaciones importantes. Estaciones ubicadas hace medio siglo en las cercanías de una ciudad y que hoy se encuentran dentro de la ciudad, no son aconsejables. Pocas son las estaciones pluviométricas que cumplen con estos requisitos en el país.

- Es probable, que exista el acoplamiento entre el ciclo solar y las precipitaciones en algunas otras zonas de ladera al interior de los Andes colombianos, donde su régimen pluviométrico anual sea de doble onda, característico del ecuador climático, como lo son algunas áreas cafeteras del Huila, Tolima, Cundinamarca y Valle.

- No es difícil encontrar en la literatura científica mundial artículos que dan cuenta de lugares del globo con influencia del ciclo de las manchas solares en la precipitación, correlación positiva y negativa, y de lugares en los que no se detecta esta relación [16, 17, 18, 19]. Lo que contrasta con la casi inexistencia de artículos colombianos que toquen la temática, y menos aún se cuenta con estudios que hayan comprobado con datos pluviométricos nacionales esa influencia y un comportamiento cíclico alrededor de los 11 años en las precipitaciones locales.

- Al menos para la hidrología regional del Eje Cafetero, no sabemos si para algunas otras zonas de Colombia, el detectarse el acoplamiento del ciclo solar en la dinámica de las precipitaciones es de suma importancia, puesto que posibilita construir modelos de pronóstico de largo plazo. El ciclo solar es absolutamente predecible por su regularidad marcada, lo que permite con una alta certidumbre su pronóstico a mediano y largo y por ende, también, se puede pronosticar su efecto pluviométrico [20]. No sucede así con las recurrencias del ENSO que son erráticas o caóticas. Claro está, que el ciclo solar es solo una causal de variabilidad climática regional, además del ENSO, existen: la ZITC, la QBO, el calentamiento global, los cambios en el uso del suelo, la corriente superficial del Chocó y varias otras más.

- Los esfuerzos por aplicar técnicas matemáticas más "robustas", en la detección de modos de variabilidad en la hidrología colombiana, que los periodogramas de densidad espectral calculados con la transformada rápida de Fourier, como el método de la transformada Hilbert-Huang basado en la descomposición de modos empíricos, tienen gran importancia [21]. Pero aquí se ha demostrado que los métodos de la estadística clásica como: correlaciones, auto-correlación, pruebas de diferencia de medias y técnicas descriptivas sencillas como el semivariograma, son capaces de dar cuenta de modos de variabilidad climática, hasta ahora no bien identificados, y de medir su efecto. Ha sido elusiva a la compresión científica la relación entre la actividad solar y el clima en la tierra, se hace necesario recurrir a las distintas técnicas matemáticas disponibles para establecer los efectos, cuando los hay, de esa relación en distintos lugares del globo. Parafraseando un adagio popular; "es preciso la combinación de todas las formas de… análisis".